\documentclass[aps,pre,floats, twocolumn,showpacs,superscriptaddress]{revtex4-1}
\usepackage{graphicx,epsfig}
\usepackage{graphics,dcolumn,bm,epic, eepic,fleqn,float}
\usepackage{amssymb,amsmath,multirow,rotate,color}
\usepackage{xcolor}
\usepackage{subfigure}
\usepackage{soul}
\usepackage{multirow}
\usepackage{colortbl}

\definecolor{CollGreen}{rgb}{0.564, 0.93,0.56}
\definecolor{SpatialBlue}{rgb}{0.39,0.58,0.93}
\definecolor{InfoCyan}{rgb}{0,1,1}
\definecolor{SocTurq}{rgb}{0.69, 0.93, 0.93}
\definecolor{TechPurple}{rgb}{1.0, 0.51, 0.98}
\definecolor{AirYellow}{rgb}{1.0, 0.964, 0.56}

\newcommand{\avg}[1]{\langle #1 \rangle}

\def\Erdos{Erd\"os}

\begin{document}

\title{Characteristic exponents of complex networks}

\author{Vincenzo Nicosia}
\affiliation{School of Mathematical Sciences, Queen Mary University of
  London, Mile End Road, E1 4NS, London (UK)}

\author{Manlio De Domenico} 
\affiliation{Departament d'Enginyeria Inform\'atica i Matem\'atiques,
  Universitat Rovira I Virgili, 43007 Tarragona, Spain}

\author{Vito Latora}
\affiliation{School of Mathematical Sciences, Queen Mary University of
  London, Mile End Road, E1 4NS, London (UK)}

\begin{abstract}
We present a novel way to characterize the structure of complex
networks by studying the statistical properties of the trajectories of
random walks over them. We consider time series corresponding to
different properties of the nodes visited by the walkers. We show that
the analysis of the fluctuations of these time series allows to define
a set of characteristic exponents which capture the local and global
organization of a network. This approach provides a way of solving two
classical problems in network science, namely the systematic
classification of networks, and the identification of the salient
properties of growing networks. The results contribute to the
construction of a unifying framework for the investigation of the
structure and dynamics of complex systems.
\end{abstract}


\maketitle

Networks are the fabric of complex systems, and network science has
provided a deeper understanding of the basic mechanisms underlying the
functioning and the evolution of diverse biological, technological and
social systems, from the human brain to the
Internet~\cite{Strogatz2001,Barabasi2002rev,Newman2003rev,
  Boccaletti2006,Barrat2008, Newman2010}. Recently, networks have been
successfully employed for the study of dynamical systems. The basic
idea consists into transforming a time series into a graph, by means
of state-space proximity and
recurrence~\cite{Zhang2006,Xu2008,Donner2010a}, transition
probabilities~\cite{Shirazi2009,campanharo11} or visibility
relationships~\cite{Lacasa2008}, and then inferring information about
the time series from the analysis of the corresponding network. These
studies have revealed the existence of intimate connections between
the statistical properties of a time series and the topology of the
network constructed from
it~\cite{Luque2009,Lacasa2010,Donner2011,Nunez2012}. However, apart
from a few
exceptions~\cite{tadic04,campanharo11,Shimada2012,Lacasa2013}, little
attention has been devoted to the dual problem, i.e. {\em studying the
  structure of complex networks by analyzing time series associated to
  them}.

In this Letter we aim at bridging this gap, by showing that a standard
analysis of the statistical properties of time series constructed from
random walks on graphs allows to characterize the topology of complex
networks. In particular, the study of fluctuations in time series
corresponding to different node properties, such as the degree, the
average degree of nearest neighbours and the clustering coefficient,
can reveal the existence of local and global correlations in the
underlying graph. In this way it is possible to associate to each
network a set of \textit{characteristic exponents} which describe the
scaling of fluctuations of each node property and capture the
intrinsic complexity of a graph in a concise way. We show that these
exponents can be employed to check the stability of the structure of
growing networks and also allow to construct a taxonomy of networks,
thus providing a quantitative, effective way of discriminating social
from biological and technological systems by looking only at their
structural properties.

\section{Model} Let $G(V,E)$ be a connected undirected graph
consisting of $N=|V|$ nodes and $K=|E|$ edges, and denote by
$A=\{a_{ij}\}$ the adjacency matrix of $G$, whose entry $a_{ij}=1$ if
there is an edge between node $i$ and node $j$, while $a_{ij}=0$
otherwise. Let us consider a random walk on $G$ described by a
time-invariant transition matrix $\Pi \equiv \{\pi_{ji}\}$.  At each
time step, a walker moves from the current node $i$ to node $j$ with a
probability $\pi_{ji}$.  The probabilities $ \pi_{ji} $ satisfy the
normalization condition $\sum_{j}\pi_{ji} = 1 \> \forall i$.

According to this definition, a walk on $G$ corresponds
to a discrete time-invariant Markov chain defined by the transition
matrix $\Pi$ on the state space $V$.
Let us now consider an instance $W$ of the walk defined by $\Pi$ on
$G$, and a real-valued property of node $i$, $\mathcal{H}_i$.  If we
indicate as $(i_0, i_1, i_2, \ldots)$ the sequence of nodes visited by
$W$, we can construct the time series $(\mathcal{H}_{i_0},
\mathcal{H}_{i_1}, \mathcal{H}_{i_2},\ldots)$.  For instance, if
$\mathcal{H}_i \equiv k_i = \sum_j a_{ij}$, we get the time series
$(k_{i_0}, k_{i_1}, k_{i_2}, \ldots)$ of the degrees of the visited
nodes.  
\begin{figure}[!t]
  \begin{center}
    \includegraphics[width=3in]{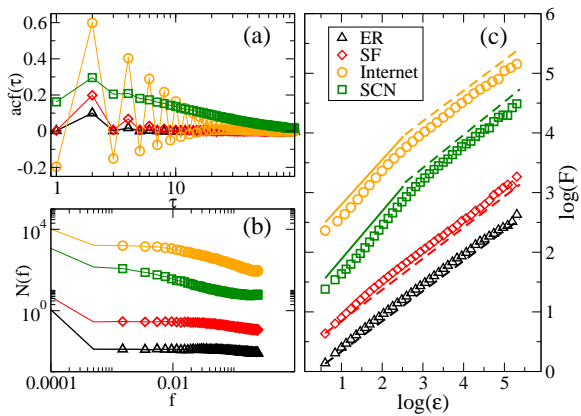}
  \end{center}
  \caption{(a) Autocorrelation function and (b) power spectrum of the
    time series constructed from the degree of nodes visited by random
    walks on an \Erdos-Renyi random graph (ER), a scale-free graph
    with $\gamma=3.0$ (SF), the Internet at the level of Autonomous
    Systems (Internet) and the collaboration network of scientists in
    condensed matter (SCN).  Panel (c): the DFA of the time series
    reveals two distinct scaling regimes in the Internet and in SCN,
    while the fluctuations in ER and SF networks are compatible with
    Gaussian noise. The plots in (b) and (c) have been vertically
    displaced to improve readability. }
  \label{fig:fig1}
\end{figure}

We denote such time series as $T(W,\mathcal{H})$, because it
depends on the node property $\mathcal{H}$, and on the specific order
in which the nodes are visited by the walk.  However, if the walk
defined by the transition matrix $\Pi$ on $G$ is irreducible, then the
topology of $G$ completely determines which sequences of values can be
produced by the walk and with which frequency~\cite{Cover1991}. Hence,
any two time series $T'(W', \mathcal{H})$ and $T''(W'', \mathcal{H})$
constructed from two walkers $W'$ and $W''$ on $G$ corresponding to
the same walk rule $\Pi$ and the same node property $\mathcal{H}$ will
have, for $t\rightarrow \infty$, the same statistical properties, and
will carry the same information about the structure of $G$. We can
therefore indicate any time series produced by a transition matrix
$\Pi$ and by node property $\mathcal{H}$ as $T(\Pi,\mathcal{H})$. We
will now show that the analysis of the time series produced by
different node properties $\mathcal{H}$ can provide useful insights on
the microscopic structure of a complex network and about its overall
organization.
We focus on the case of classical random walks, i.e. we set
$  \pi_{ji} = a_{ij} / k_i$. 
%
Notice that with this rule the walkers visit each edge of a connected graph 
$G$ with uniform probability,
so that the time series constructed from random walks on $G$ contain
information about the distribution and correlations of the chosen node
property $\mathcal{H}$ throughout the network.
We consider three possible choices of $\mathcal{H}$, namely the
node degree, $\mathcal{H}_i \equiv k_i$, the average degree of first
neighbours of a node, $\mathcal{H}_i \equiv
k^{nn}_i=k_i^{-1}\sum_{j}a_{ij}k_j$, and the node clustering coefficient, 
$\mathcal{H}_i \equiv C_i$, where $C_i$ is the number of
closed triads centered on $i$ divided by the total possible number
$k_i(k_i-1)/2$ of such triads. We decided to focus on these three
node properties because broad-tailed degree distributions ($P(k)\sim
k^{-\gamma}$, $2<\gamma<3$), the presence of non-trivial degree
correlations ($k^{nn}(i) \sim k_i^{\nu}$) and the abundance of
triangles ($\avg{C_i} \gg 0$) are the basic features of
most complex networks~\cite{Barabasi2002rev,Newman2003rev,
  Boccaletti2006}.

\begin{figure}[!b]
  \begin{center}
    \includegraphics[width=3in]{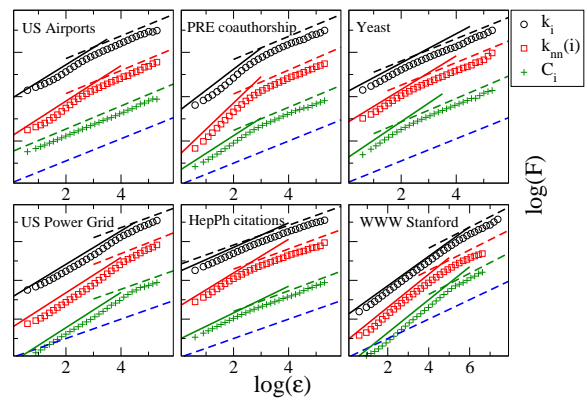}
  \end{center}
  \caption{DFA of different node properties.  We considered time
    series produced by $k_i$ , $k^{nn}_i$ and $C_{i}$, and six
    networks, namely the US airports network~\cite{Colizza2007}, the
    co-authorship network from papers published in Physical Review E,
    the budding yeast protein interaction network~\cite{Sun2003}, the
    US power grid~\cite{Watts1998}, a citation network in high-energy
    physics~\cite{Gehrke2003} and the World Wide
    Web~\cite{Leskovec2009}. The plots have been vertically displaced
    to enhance readability. We observe two scaling regimes, with the
    actual values of the two characteristic exponents $\nu_1$ and
    $\nu_2$ varying across different networks. The dashed blue line in
    each panel is the DFA of the time series of the corresponding
    randomized networks ($F(\varepsilon) \sim \varepsilon^{1/2}$),
    averaged over 1000 realizations.}
  \label{fig:fig2}
\end{figure}

\begin{figure*}[!t]
  \begin{center}
    \subfigure[]{
      \includegraphics[width=3.1in]{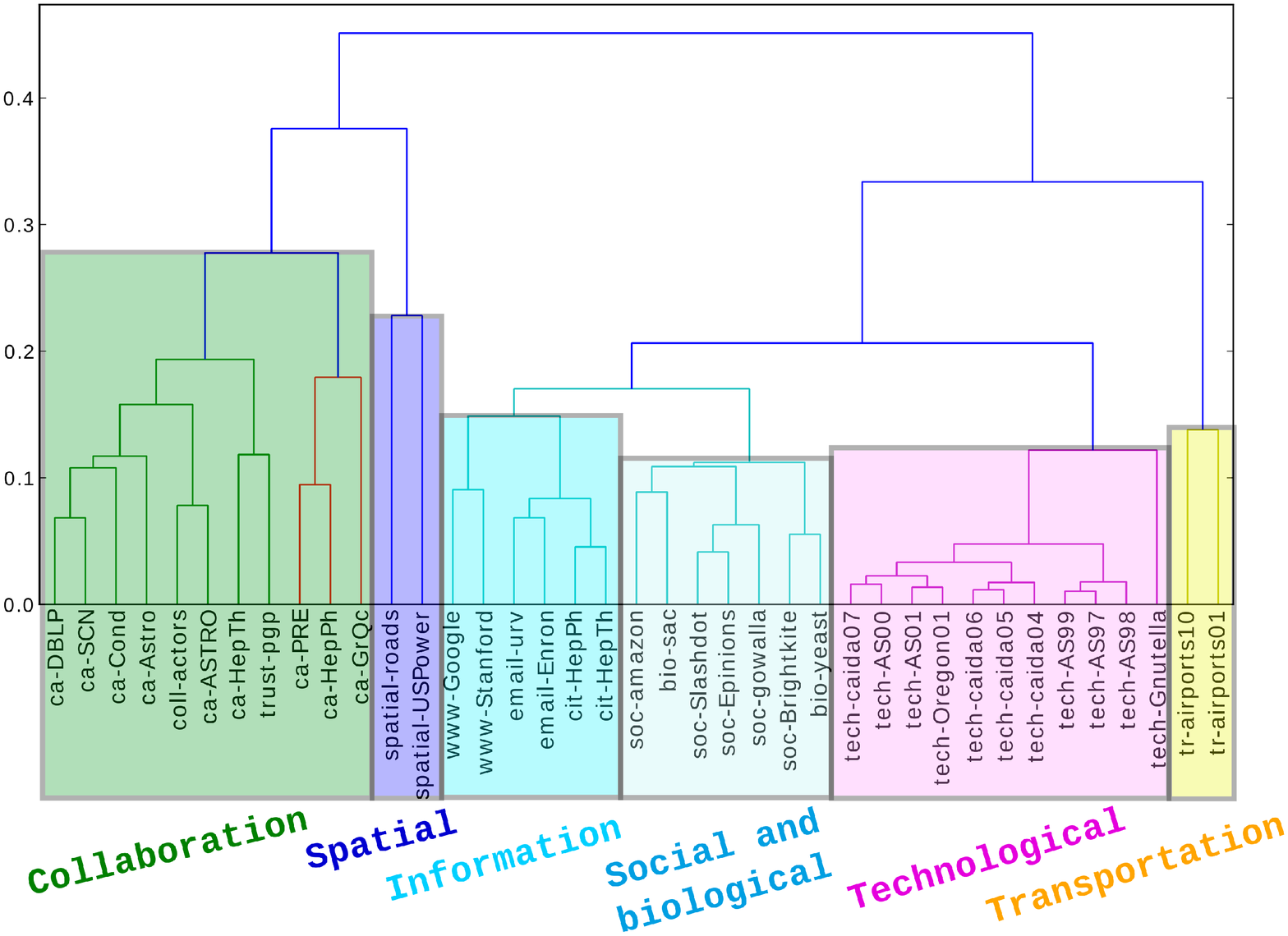}
    }
    \subfigure[]{
      \includegraphics[width=3.1in]{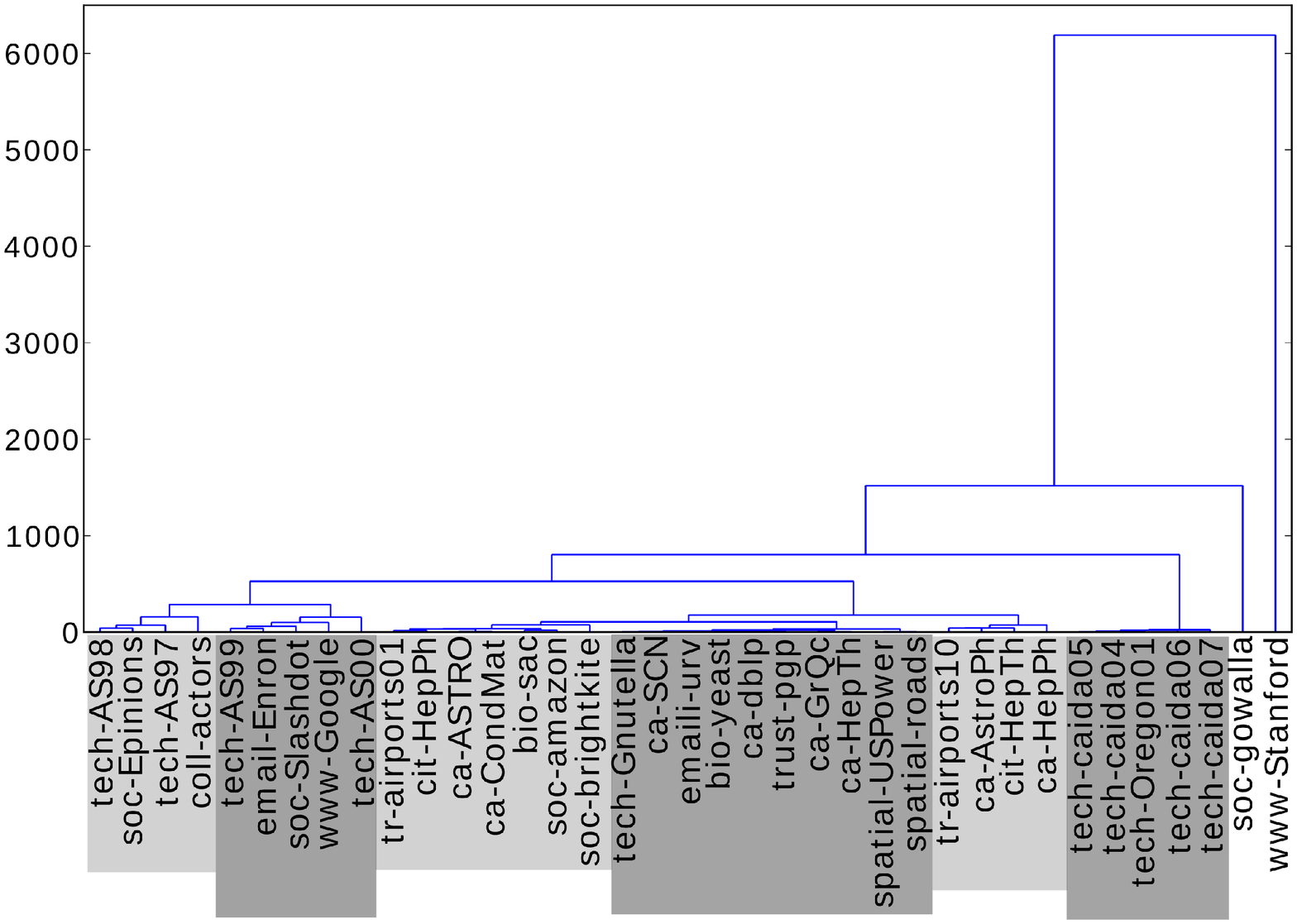}
    }
  \end{center}
  \caption{(a) The dendrogram represents the hierarchical clustering
    of 39 real-world networks obtained by using the characteristic
    exponents, $\nu_1$ and $\nu_2$, of the three time series
    (respectively based on $k_i$, $k^{nn}_i$ and $C_i$).  Notice the
    presence of well-defined, meaningful clusters of networks, namely
    collaboration and trust networks (green), spatially-embedded
    networks (blue), information networks (bright cyan), biological
    and online social networks (dark cyan), technological
    infrastructure networks (purple), and air transportation networks
    (yellow). (b) For comparison, we show the results of hierarchical
    clustering based on the mean and standard deviations of $k_i$,
    $k^{nn}_i$ and $C_i$, in which the clusters always contain
    networks of different nature.}
  \label{fig:fig3}
\end{figure*}

\section{Results} In fig.~\ref{fig:fig1}(a) and \ref{fig:fig1}(b)
we respectively report the autocorrelation function (ACF) and the
power spectrum (PS) of the degree-based time series
($\mathcal{H}_i=k_i$) obtained in an \Erdos-Renyi random graph (ER), a
scale-free graph (SF) constructed by the configuration model
\cite{Newman2010}, and two real-world complex networks, namely the
Internet at the level of Autonomous Systems
(Internet)~\cite{Pastor-Satorras2001} and the network of co-authorship
in condensed matter (SCN)~\cite{Newman2001}.  As expected, the ACF of
ER and SF decays pretty fast and the corresponding PS is almost flat,
indicating the absence of degree correlations.  Conversely, the
degree-based time series obtained from real-world networks exhibit
broad tails both in the ACF and the in the PS, a clear indication of
the presence of long-range degree correlations. The peaks at even
values of $\tau$ in the ACF of Internet are due to the presence of
disassortative degree correlations~\cite{Note1}.
An iterative surrogate analysis~\cite{Schreiber1996} has also
confirmed that these time series are highly non-linear.  A
non-parametric statistical test~\cite{dedomenico2010fast}, not
depending on delay embedding reconstruction, suggested that such time
series are non-linear or non-stationary with high confidence level.

In the following we report the results of the multifractal Detrended
Fluctuation Analysis (DFA)~\cite{kantelhardt2002multifractal}, a
standard non-linear analysis technique which allows to detect the
presence of long-range correlations and to quantify the self-affinity
of a time series, even if generated by a non-stationary process.
Given a time series $(\mathcal{H}_{i_0}, \mathcal{H}_{i_1},
\mathcal{H}_{i_2},\ldots)$ we consider $\ell$ time-windows of length
$\varepsilon$; then, we remove the local linear trend in each
time-window to obtain the detrended time series
$(\overline{\mathcal{H}}_{i_0}, \overline{\mathcal{H}}_{i_1},
\overline{\mathcal{H}}_{i_2},\ldots)$ and we compute the local
variance $\sigma^{2}(\ell,\varepsilon)$ of the detrended
fluctuations. We evaluate the structure function $F(\varepsilon)$ by
averaging $\sigma^{2}(\ell,\varepsilon)$ over all time-windows whose
length is equal to $\varepsilon$, and we plot $F(\varepsilon)$ as a
function of $\varepsilon$. The procedure can be generalized to build a
set of structure functions depending on a parameter
$q$~\cite{Hurst51,Hurst56,heneghan2000establishing}, but here we focus
on $q=2$, allowing a physical interpretation of the results in term of
diffusivity.  

If the graph $G$ is $\mathcal{H}$-uncorrelated, i.e. if
the probability to find the edge $(i,j)$ connecting node $i$ to node
$j$ does not depend on the values $\mathcal{H}_i$ and $\mathcal{H}_j$,
then the fluctuations of the corresponding time series
$T(\Pi,\mathcal{H})$ obtained from a random walk on $G$ will be
indistinguishable from an uncorrelated Gaussian noise, for which we
have $F(\varepsilon) \sim {\varepsilon}^{1/2}$. Conversely, a scaling
behaviour $F(\varepsilon)\sim \varepsilon^{\alpha}$ with $\alpha\neq
1/2$ is a clear signal of the existence of $\mathcal{H}$-correlations
in the original graph $G$, and the value of $\alpha$ is a proxy for
the magnitude of such correlations.

In fig.~\ref{fig:fig1}(c) we report the results of the DFA of
$\mathcal{H}_i\equiv k_i$ for the same four networks considered in
panel (a) and (b).  As expected, degree fluctuations in ER and SF are
compatible with Gaussian noise ($F(\varepsilon)\sim
\varepsilon^{1/2}$), since the node degrees in these networks are
uncorrelated. Conversely, $F(\varepsilon)$ plots corresponding to time
series generated by walkers on the Internet and on the SCN appreciably
deviate from Gaussian noise and are characterized by two different
regimes~\footnote{To observe the two scaling regimes, it is necessary
  to analyse sufficiently long time-series. In particular, for a
  network with $N$ nodes we suggest to generate time-series of length
  at least equal to $10^2\times N$, or better $\sim 10^3\times N$, in
  order to guarantee that each edge of the graph has been traversed a
  sufficiently high number of times.}.  In the first regime,
corresponding to small values of $\varepsilon$, both time series are
super-diffusive, i.e. $F(\varepsilon)\sim \varepsilon^{\nu_1}$ with
$\nu_1>1/2$ ($\nu_1\simeq 0.75$ for Internet and $\nu_1\simeq 0.80$
for SCN), while for large values of $\varepsilon$ their behaviour is
almost Gaussian ($F(\varepsilon)\sim \varepsilon^{\nu_2}$ with
$\nu_2\simeq 0.51 $ for Internet and $\nu_2\simeq 0.52$ for SCN).
In fig.~\ref{fig:fig2} we report the results of the DFA of time series
generated by $\mathcal{H}_i= k_i$, $\mathcal{H}_i= k^{nn}_i$ and
$\mathcal{H}_i=C_i$ in six real-world networks of different
nature~\cite{Note2}. The same two-regime behavior shown in
fig.~\ref{fig:fig1} for degree-based time series, is also found for
the time series generated by $k^{nn}_i$ and $C_i$.  The two scaling
regimes are a signature that the networks look different, with respect
to degree, degree correlations and clustering, when observed at a
local or at a global scale. On the one hand, the super-diffusive
behaviour observed for small values of $\varepsilon$ ($F(\varepsilon)
\sim \varepsilon^{\nu_1}$) indicates that a walker which explores the
network for relatively short time intervals will observe correlated
fluctuations in the properties of the nodes it visits, a clear signal
of the presence of $\mathcal{H}$-correlations. On the other hand, the
almost-Gaussian behaviour corresponding to large values of
$\varepsilon$ ($F(\varepsilon) \sim \varepsilon^{\nu_2}$) suggests
that at a larger scale (i.e., if the walk continues for a sufficiently
long time), the network appears uncorrelated. The transition point
$\varepsilon_{c}$ that separates the two regimes corresponds to the
typical scale of $\mathcal{H}$--correlations, i.e. the typical walk
length above which local heterogeneities and correlations in the
values of $\mathcal{H}$ become less important and all the walks on the
network can be considered a homogeneous representation of the typical
$\mathcal{H}$-fluctuations of the graph. We notice that in some cases
the exponent $\nu_2$ can be substantially larger than $0.5$, like in
the case of the US power grid~\cite{Watts1998}, for which we have
$\nu_2>0.65 $ for all the three time series. In this particular case,
the super-diffusive behavior for large values of $\varepsilon$ is due
to the fact that the network is embedded in a 2D space and has a
strongly self-similar structure~\cite{Daqing2011}.

Although the presence of two scaling regimes seems to be a ubiquitous
feature of different real-world networks, independently of their
origin and nature, fig.~\ref{fig:fig2} indicates that the actual
values of the two exponents $\nu_1$ and $\nu_2$ may vary a lot for
different node properties of the same network and, more importantly,
for the same node property across different networks. In the following
we show that these scaling exponents capture some key properties of a
graph and can be employed to construct a taxonomy of
networks~\cite{Estrada2007,Onnela2012}.

We considered a data set of 39 medium-to-large sized ($N\sim 10^4$ to
$N\sim 10^{6}$) real-world networks representing different social,
biological and technological systems. We assigned to each graph $G$ a
point $p(G)\in \mathbb{R}^6$ identified by the values of the six
scaling exponents obtained from the DFA of time series of degree,
clustering coefficient and average degree of first neighbours. Then,
we performed a hierarchical clustering on the resulting set of points,
subsequently merging together at each step the two clusters whose
points were separated by the smallest distance in $\mathbb{R}^6$.
In fig.~\ref{fig:fig3}(a) we report the resulting dendrogram, where
the six large clusters identified (highlighted with different colors)
correspond to networks with different functions.  From left to right:
the green cluster contains all the co-authorship
(\cite{Newman2001,Leskovec2007a}), trust (PGP~\cite{Boguna2004}) and
collaboration networks (IMDb co-starring network~\cite{Watts1998});
the blue cluster includes spatial networks (US power
grid~\cite{Watts1998} and the Pennsylvania road
network~\cite{Leskovec2009}); the bright-cyan cluster contains
information networks, such as the WWW~\cite{Leskovec2009}, citation
networks~\cite{Gehrke2003}, and email communication
networks~\cite{Guimera2003,Leskovec2009}; the dark-cyan cluster
includes online social
networks~\cite{Richardson2003,Leskovec2009,Cho2011} and
proteomes~\cite{Watts1998,Colizza2005}; the purple cluster contains
technological networks, including snapshot of the Internet sampled at
different times by different
institutions~\cite{Pastor-Satorras2001,Leskovec2005,COSIN} and the
Gnutella peer-to-peer file-sharing
network~\cite{Ripeanu2002}. Finally, the networks of US airports at
two different times~\cite{Colizza2007} are put together in the yellow
cluster. The accuracy of characteristic exponents in classifying
networks of different nature is quite remarkable~\footnote{We also
  tried to perform hierarchical clustering of the 39 networks by
  using, for each time-series, only the exponent $\nu_1$ corresponding
  to the scaling of $F(\varepsilon)$ for small values of
  $\varepsilon$. However, the resulting classification is not as neat
  and as clear as the one reported in fig.~\ref{fig:fig3}.}, and
becomes evident by comparing the results of fig.~\ref{fig:fig3}(a)
with those of hierarchical clustering based on the mean and standard
deviations of $k_i$, $k^{nn}_i$ and $C_i$, reported in
fig.~\ref{fig:fig3}(b). While in the former case clusters represent
homogeneous groups of networks, in the latter case each cluster always
contains networks of different nature.

The results shown in fig.~\ref{fig:fig2} and fig.~\ref{fig:fig3}
suggest that the scaling exponents of the time series produced by
random walkers visiting a complex network are indeed a key feature to
characterize the network.  Hence, we name them \textit{characteristic
  exponents} of the network (Table~\ref{table:table1} reports the
characteristic exponents of all the complex networks considered in
this study).
\begin{figure}[t]
  \begin{center}
    \includegraphics[width=3.1in]{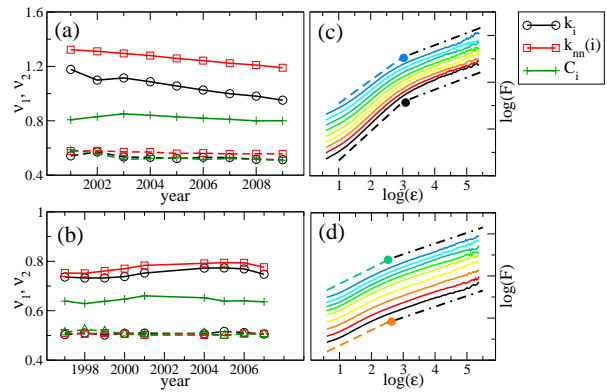}
  \end{center}
  \caption{(color online) Characteristic exponents of (a) a
    co-authorship network from Physical Review E (PRE) and (b) the
    Internet at the level of Autonomous Systems, over a period of ten
    years. In PRE the characteristic exponents for $k_i$ and
    $k^{nn}_i$ decrease over time, while the exponents for $C_i$
    remain constant. In the Internet, all the characteristic exponents
    are constant.  Panel (c) and panel (d) report the detail of the
    DFA for degree time series, respectively for PRE and Internet. The
    plots have been vertically displaced to enhance readability, with
    the topmost curve corresponding to the most recent network.}
  \label{fig:fig4}
\end{figure}

It is also interesting to investigate how the characteristic exponents
of growing graphs change over time.  In fig.~\ref{fig:fig4}(a) and
\ref{fig:fig4}(b) we show the temporal evolution of the characteristic
exponents of $k_i$, $k^{nn}_i$ and $C_i$ respectively for the
collaboration network of authors in APS Physical Review E (PRE) and
for the Internet. Both networks have grown by a factor $\sim 9$ in the
considered time intervals. However, while in PRE the characteristic
exponent $\nu_1$ for $k_i$ and $k^{nn}_i$ exhibits a clear decrease
over time, the characteristic exponents of the Internet have remained
constant in the considered 10-years interval.
The different temporal behaviour of the characteristic exponents is
probably due to the peculiar dynamics of edge formation in the two
networks.  In fact, in a co-authorship network a node continues to
accumulate edges over time, even if the majority of these edges
correspond to collaborations which are not active any more. Evidently,
the continuous addition of edges drives the network towards a
homogenization of degree and clustering correlations.
Conversely, the number of neighbours of a node in the Internet 
cannot increase indefinitely, due to technological and economical
constraints. In fact, connecting to more peers usually implies
handling more Internet traffic, which in turn requires more bandwidth
and new hardware, and translates into an economical investment. These
constraints are mostly independent from network size, thus having the
same impact on the network growth at different times. This might explain 
why the structure of correlations has remained stable over time.

\begin{figure}[!t]
  \begin{center}
    \includegraphics[width=3in]{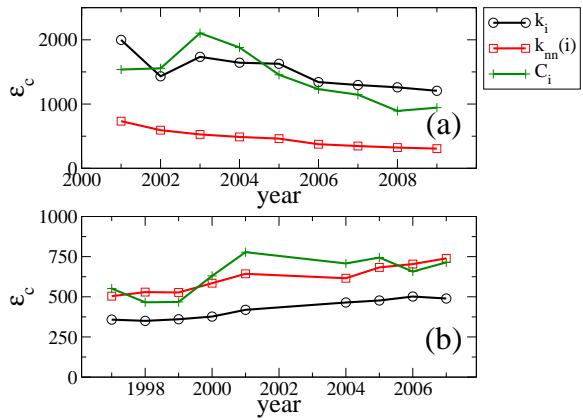}
  \end{center}
  \caption{(color online) The value $\varepsilon_c$ at which the
    scaling of the structure function $F(\varepsilon)$ changes,
    respectively for (a) PRE and (b) Internet, as a function of
    time. Despite $\varepsilon_c$ can slightly increase or decrease
    over time, its value remains of the same order of magnitude and
    corresponds, in both cases, to the visit of just a few hundred
    nodes.}
  \label{fig:fig5}
\end{figure}

\begin{table}[!t]
  \small
  \begin{tabular}{|l|ccc|ccc|}
    \hline Network & \multicolumn{3}{|c|}{$\nu_1$} &
    \multicolumn{3}{|c|}{$\nu_2$} \\ \hline
    & $k_i$ & $k^{nn}_i$ & $C_i$ & $k_i$ & $k^{nn}_i$ & $C_i$\\
    \hline
    \rowcolor{CollGreen}
    ca-DBLP & 0.838 & 1.092 & 0.769 & 0.519 & 0.523 & 0.516\\
    \rowcolor{CollGreen}
    ca-SCN & 0.803 & 1.044 & 0.757 & 0.518 & 0.554 & 0.524\\
    \rowcolor{CollGreen}
    ca-Cond & 0.726 & 1.032 & 0.748 & 0.509 & 0.564 & 0.523\\
    \rowcolor{CollGreen}
    ca-Astro & 0.739 & 1.119 & 0.810 & 0.526 & 0.588 & 0.527\\
    \rowcolor{CollGreen}
    coll-actors & 0.789 & 1.211 & 0.763 & 0.506 & 0.554 & 0.516\\
    \rowcolor{CollGreen}
    ca-ASTRO & 0.838 & 1.215 & 0.780 & 0.531 & 0.606 & 0.506\\
    \rowcolor{CollGreen}
    ca-HepTh & 0.811 & 1.034 & 0.831 & 0.535 & 0.601 & 0.615\\
    \rowcolor{CollGreen}
    trust-pgp & 0.853 & 1.035 & 0.901 & 0.515 & 0.552 & 0.547\\
    \rowcolor{CollGreen}
    ca-PRE & 0.953 & 1.188 & 0.799 & 0.522 & 0.556 & 0.520\\
    \rowcolor{CollGreen}
    ca-HepPh & 0.993 & 1.220 & 0.860 & 0.549 & 0.585 & 0.554\\
    \rowcolor{CollGreen}
    ca-GrQc & 1.006 & 1.254 & 0.924 & 0.610 & 0.672 & 0.596\\
    \rowcolor{SpatialBlue}
    spatial-roads & 0.808 & 0.833 & 0.839 & 0.602 & 0.607 & 0.740\\
    \rowcolor{SpatialBlue}
    spatial-USPower & 0.863 & 0.964 & 0.890 & 0.656 & 0.752 & 0.685\\
    \rowcolor{InfoCyan}
    www-Google & 0.695 & 0.874 & 0.802 & 0.510 & 0.512 & 0.543\\
    \rowcolor{InfoCyan}
    www-Stanford & 0.760 & 0.854 & 0.850 & 0.542 & 0.528 & 0.548\\
    \rowcolor{InfoCyan}
    email-urv & 0.662 & 0.876 & 0.743 & 0.501 & 0.526 & 0.514\\
    \rowcolor{InfoCyan}
    email-Enron & 0.681 & 0.813 & 0.735 & 0.511 & 0.516 & 0.522\\
    \rowcolor{InfoCyan}
    cit-HepPh & 0.629 & 0.871 & 0.696 & 0.506 & 0.524 & 0.518\\
    \rowcolor{InfoCyan}
    cit-HepTh & 0.593 & 0.847 & 0.700 & 0.515 & 0.518 & 0.521\\
    \rowcolor{SocTurq}
    soc-amazon & 0.582 & 0.708 & 0.865 & 0.516 & 0.547 & 0.525\\
    \rowcolor{SocTurq}
    bio-sac & 0.648 & 0.668 & 0.858 & 0.508 & 0.504 & 0.531\\
    \rowcolor{SocTurq}
    soc-Slashdot & 0.604 & 0.747 & 0.757 & 0.493 & 0.511 & 0.512\\
    \rowcolor{SocTurq}
    soc-Epinions & 0.596 & 0.713 & 0.757 & 0.514 & 0.509 & 0.519\\
    \rowcolor{SocTurq}
    soc-gowalla & 0.630 & 0.688 & 0.782 & 0.502 & 0.523 & 0.528\\
    \rowcolor{SocTurq}
    soc-Brightkite & 0.639 & 0.801 & 0.828 & 0.509 & 0.536 & 0.526\\
    \rowcolor{SocTurq}
    bio-yeast & 0.663 & 0.758 & 0.837 & 0.504 & 0.513 & 0.530\\
    \rowcolor{TechPurple}
    tech-caida07 & 0.748 & 0.776 & 0.636 & 0.506 & 0.507 & 0.506\\
    \rowcolor{TechPurple}
    tech-AS00 & 0.738 & 0.770 & 0.647 & 0.509 & 0.505 & 0.508\\
    \rowcolor{TechPurple}
    tech-AS01 & 0.752 & 0.783 & 0.660 & 0.509 & 0.501 & 0.504\\
    \rowcolor{TechPurple}
    tech-Oregon01 & 0.748 & 0.781 & 0.655 & 0.503 & 0.510 & 0.502\\
    \rowcolor{TechPurple}
    tech-caida06 & 0.769 & 0.794 & 0.640 & 0.512 & 0.507 & 0.507\\
    \rowcolor{TechPurple}
    tech-caida05 & 0.773 & 0.794 & 0.639 & 0.516 & 0.501 & 0.499\\
    \rowcolor{TechPurple}
    tech-caida04 & 0.772 & 0.792 & 0.652 & 0.507 & 0.502 & 0.510\\
    \rowcolor{TechPurple}
    tech-AS99 & 0.732 & 0.761 & 0.638 & 0.502 & 0.510 & 0.517\\
    \rowcolor{TechPurple}
    tech-AS97 & 0.736 & 0.752 & 0.639 & 0.503 & 0.509 & 0.512\\
    \rowcolor{TechPurple}
    tech-AS98 & 0.732 & 0.751 & 0.629 & 0.510 & 0.506 & 0.524\\
    \rowcolor{TechPurple}
    tech-Gnutella & 0.640 & 0.714 & 0.642 & 0.501 & 0.505 & 0.502\\
    \rowcolor{AirYellow}
    tr-airports01 & 0.770 & 0.926 & 0.516 & 0.518 & 0.542 & 0.518\\
    \rowcolor{AirYellow}
    tr-airports10 & 0.866 & 1.001 & 0.519 & 0.473 & 0.494 & 0.519\\
    \hline
  \end{tabular}
  \caption{The two characteristic exponents $\nu_1$ and $\nu_2$ of
    time series constructed from node degree ($k_i$), average degree
    of first neighbours ($k^{nn}_i$) and node clustering coefficient
    ($C_i$) of real-world complex networks. The color correspond to
    the class to which a network belongs, i.e. coauthorship,
    collaboration and trust networks (green), spatial networks (blue),
    information and citation networks (bright cyan), social networks
    and proteomes (dark cyan), technological networks (purple) and air
    transportation networks (yellow). }
  \label{table:table1}
\end{table}

Finally, we check whether the position $\varepsilon_{c}$ of the
cut-off of the structure function does depend on the size of the
graph, and to which extent. To this aim, we show in
fig.~\ref{fig:fig5} the approximate value of $\varepsilon_c$ for the
PRE and Internet networks, as a function of time. Notice that as the
networks grow the corresponding values of $\varepsilon_c$ change
slightly for all the time-series, but we observe opposite trends in
the two cases. In particular, $\varepsilon_c$ usually decreases for
PRE and increases in Internet. This means that $\varepsilon_c$ is not
simply determined by the size of the network (otherwise we should have
observed a similar behaviour in both networks), but is instead
intimately related to the local organization of the graph.

It is also worth noticing that, despite the presence of these trends,
$\varepsilon_c$ usually remains of the same order while both networks
have grown by an order of magnitude in the considered time
intervals. For instance, in the time-series of degrees of PRE [see
  fig.~\ref{fig:fig5}(a)] $\varepsilon_c$ remains in the range
$[1250:1750]$, while for the time-series of $k^{nn}_i$ it is in the
range $[300:600]$. If we take into account the fact that a random walk
on any of the snapshots of the PRE collaboration network typically
requires $O(10^6)$ time-steps in order to visit all the nodes at least
once, and that this network has strong communities and a high value of
clustering coefficient (and both these factors contribute to keep a
walker confined on a small set of nodes), then we realise that
$\varepsilon_c\sim 10^3$ corresponds indeed to the exploration of a
relatively small region of the graph, which usually includes no more
than a few hundred nodes. Similarly, in the Internet network
[fig.~\ref{fig:fig5}(b)] $\varepsilon_c$ is in the interval
$[350:500]$ for $k_i$ and in $[450:700]$ for $k^{nn}_i$ and $C_i$,
which again correspond to visiting a relatively small portion of the
graph. These results suggest that the values of the cut-off in the
scaling of the structure functions tend to remain practically stable
over time, even when the network undergoes substantial expansion.

Summing up, in this work we reported on the discovery of an intimate
connection between the structure of a network, the properties of
time-series extracted from it, and the capability of such time-series
to carry useful information about the overall organization of the
network. We have shown that the characteristic exponents corresponding
to degree, node clustering coefficient and average degree of first
neighbours can be used to cluster networks, and to distinguish social,
collaboration, biological, information, transportation and spatial
networks only by looking at their structure. The procedure described
in this work is quite general, and can be used to extract
characteristic exponents corresponding to any desired node or link
property, thus allowing for a finer and more accurate classification
of complex networks.

\acknowledgments
VN and VL acknowledge support from the EU-LASAGNE Project, Contract
No.318132 (STREP) funded by the European Commission. MDD is supported
by the FET-Proactive project PLEXMATH (FP7-ICT-2011-8; grant number
317614) and MULTIPLEX (317532) funded by the European Commission.

\end{document}